\documentclass[twocolumn,preprintnumbers,amsmath,amssymb]{revtex4-2}
\usepackage{graphicx}  
\usepackage{sidecap}
\begin{document}
%
\title{Terahertz bolometric detectors  based on graphene field-effect transistors with the composite h-BN/black-P/h-BN gate layers using  plasmonic resonances
}
\author{ M.~Ryzhii$^{1}$, V.~Ryzhii$^{2}$,  M.~S.~Shur$^3$,
V.~Mitin$^4$,  C.~Tang$^{2,5}$, and T.~Otsuji$^2$ }
\address{
$^1$Department of Computer Science and Engineering, University of Aizu, Aizu-Wakamatsu 965-8580, Japan\\
$^2$Research Institute of Electrical Communication,~Tohoku University,~Sendai~ 980-8577, 
Japan\\
$^3$Department of Electrical,~Computer,~and~Systems~Engineering, Rensselaer Polytechnic Institute,~Troy,~New York~12180,~USA\\
$^4$Department~of~Electrical Engineering,~University at Buffalo, SUNY, Buffalo, New York 14260 USA\\
$^5$Frontier Research Institute for Interdisciplinary Sciences,~Tohoku University,~Sendai~ 980-8578, Japan  
}
\begin{abstract}
We propose and analyze the performance of  terahertz (THz) room-temperature bolometric detectors based on the graphene channel
field-effect transistors (GC-FET). 
These detectors comprise the gate barrier layer (BL) composed of the lateral
hexagonal-Boron Nitride black-Phosphorus/ hexagonal-Boron Nitride (h-BN/b-P/h-BN) structure.
The main part of the GC is encapsulated in h-BN, whereas a short section of the GC is sandwiched between the b-P gate BL and the h-BN bottom layer. 
The b-P gate BL serves as the window for the electron thermionic current from the GC. 
The electron mobility in  the GC section  encapsulated in h-BN can be fairly large. 
This might  enable a strong resonant plasmonic response of the GC-FET detectors despite relatively lower electron mobility in the GC section covered by the b-P window BL. 
The narrow b-P window diminishes the Peltier cooling and enhances the detector performance.
The proposed device structure and its operation principle promote elevated values of the room-temperature GC-FET THz detector responsivity and other characteristics, especially at the plasmonic resonances.  
\end{abstract}

%
\maketitle
%


\section{Introduction}

The favorable band alignment of the graphene channel (GC) and the 
black-Arsenic$_{x}$Phosphorus$_{1-x}$ (b-AsP) layers~\cite{1,2,3,4}
enables new opportunities for electron and optoelectronic device applications.
In particular, we recently  proposed the terahertz (THz) bolometric detectors based on the field-effect transistors (FETs) with the b-AsP gate barrier layer (BL) and the metal gate (MG)~\cite{5,6}. 
The operation of such bolometric detectors is associated with the heating of the two-dimensional electron (hole) system (2DEG) by the THz radiation resulting in the reinforcement
of the thermionic electron emission from the GC to the MG via the b-AsP BL.
The resonant excitation of the plasmonic oscillations  in the GC~\cite{7,8,9}
can result in a substantial rise  of the electron heating efficiency and can lead to improvement of the detector characteristics, in particular, enhancement of the detector responsivity.
However, the plasmonic resonances are sensitive to electron scattering, limiting the sharpness of the resonances and the responsivity peaks. 
This problem is linked to the prospects of elevating the room temperature mobility of the GCs contacted with the b-AsP BLs.

In this paper, we propose and evaluate the room-temperature THz bolometric detectors based on the GC-FETs, in which the thermionic electron current passes through a relatively narrow b-P collector window in the primarily h-BN gate BL. 
The main part of the GC/BL interface is made of h-BN, so the GC exhibits higher electron mobility in comparison with the GC covered by the b-P gate BL~\cite{10,11,12,13,14,15}.
Using the composite h-BN/b-AsP/h-BN gate BL, one can improve the GC-FET detector performance due to:\\
(i) a substantial decrease in the plasmonic oscillations damping;\\
(ii) a weaker hot electron energy flux from the GL via the b-P collector window resulting in the diminishing of the Peltier cooling of the 2DEG~\cite{16} because of a smaller length of the b-P section of the BL.

As shown in the following, these features of the GC-FETs with the composite gate BL enable a marked enhancement of the detector performance.

\begin{figure*}[t]
\centering
\includegraphics[width=17.0cm]{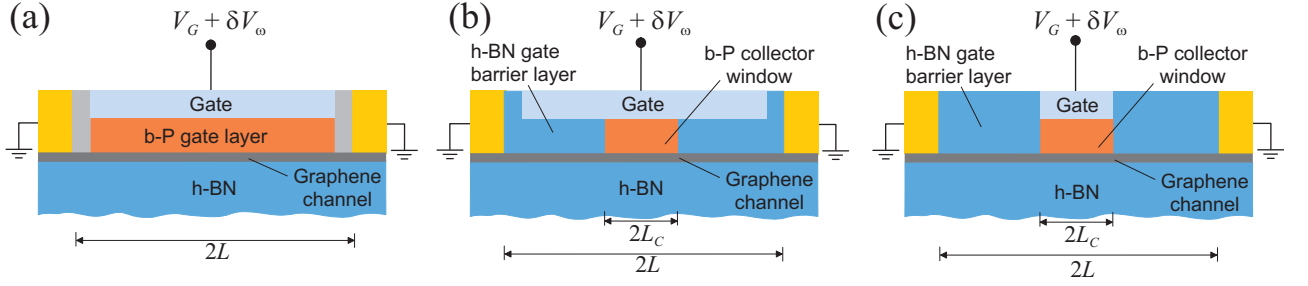}
\caption{Schematic views of  the GC-FET bolometric detector structures with  (a) the b-P gate BL~\cite{6,15},
(b) the composite h-BN/black-AsP/h-BN gate BL, and (c) the composite  gate BL 
but with shortened MG.}
\label{Fig1}
\end{figure*}

\begin{figure}[t]
\centering
\includegraphics[width=8.0cm]{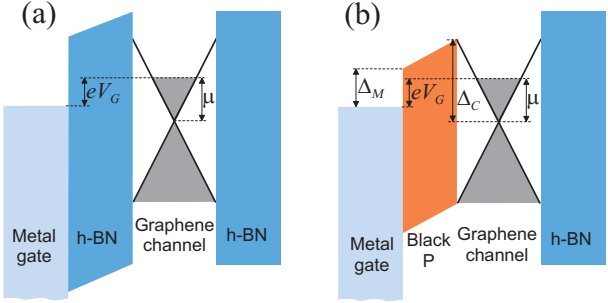}
\caption{The band diagrams of the GC-FET with the composite gate BL [shown in Fig.~1(b)] at its different cross-sections (a) $
L_C < |x| < L$, i.e., in the region when the gate BL consists of the h-BN
and (b)  $|x| < L_C$ (in the collector window).}
\label{Fig2}
\end{figure}

\section{GC-FET bolometric detector device structure }

Figure 1 shows the structures of the GC-FET bolometric detectors with different BLs and MGs. 
The structure of the GC-FET with the uniform b-P gate BL [see Fig.~1(a)] was considered previously~\cite{8,15}. 
Here, we focus
on the GC-FET bolometric detectors operating at room temperature with the composite BLs, schematically shown in Fig.~1(b).
The devices with the structures seen in Fig.~1(c)  are briefly discussed in the Comments section.
The structure of the GC-FET bolometric detectors under consideration [see Fig.~1(b)] comprises the n-doped GC sandwiched between the substrate (h-BN) and the h-BN gate BL  with a narrow b-P window. 
This structure is covered by the MG and supplied by  the side contacts (corresponding to the FET source and drain).
The length of the collector window $2L_C$ is smaller than the length of the GC $2L$. 
Figure~2 shows these GC-FET bolometric detectors  band diagram (in the direction $z$ perpendicular to the layers).
The relatively low energy barrier for the electrons in the GL enables thermionic emission (in the region $|x| < L_C$) above the energy barrier at the GC/b-P interface. 
In contrast, this energy barrier in the regions $L_C <|x|<L$  is fairly high, preventing electron emission from this section of the GC.

We assume that the sufficiently high GC doping level  provides at the bias voltage  $V_G =0$ the flat-band condition
$\Delta_M = \Delta_C - \mu_D$. 
Here
 $\Delta_M$ and $\Delta_C$ are the pertinent band offsets,  $\mu_D = \hbar\,v_W\sqrt{\pi\Sigma_D}$ is the electron Fermi energy in the GC, where
$v_W \simeq 10^8$~cm/s is the electron velocity in GCs,
$\Sigma_D$ is the donor density in the GC (or the density of the remote donors in the substrate near the GC), and $\hbar$ is the Planck constant.
In the GC-FETs under consideration with the Al MG, $\Delta_M = 85$~meV and $\Delta_C = 225$~meV, so that $\mu_D = 140$~meV, as shown in Table I. 
The latter occurs at $\Sigma_D \simeq 1.6\times10^{12}$~cm$^{-2}$ (see Ref.~\cite{8} and the references therein). 
When $V_G > 0$, the bottom of the b-P BL conduction band is somewhat inclined, as shown in Fig.~2(b).
Below we consider the GC-FETs with not too thin BLs in the range of relatively low bias voltages, so that the electron tunneling through the BLs is negligible.

Apart from the bias voltage  $V_{G} \geq 0$, the signal voltage $\delta V_{\omega} \exp(-i\omega t)$  generated by the THz radiation received by an antenna is applied between the GL and the MG. 
Here $\delta V_{\omega}$ and $\omega$ are the amplitude and frequency of the THz signal.
This connection corresponds to the contact of one antenna lead with the MG and another with both the GC-FET side contacts.
The signal voltage excites the ac current along the GC, resulting in electron heating in the GC. 
The latter  stimulates thermionic emission from the GC over the b-P section of the BL.

\section{Rectified current }

The variation of the rectified thermionic gate current, $\langle \delta J_{\omega}\rangle$,  associated with the incoming THz signal (and averaged over its period and serving as 
the detector output signal) is presented  in the following form:

\begin{eqnarray}\label{eq1}
\langle \delta J_{\omega}\rangle = j^{max} F
 H\int_{-L_C}^{L_C}\,dx\frac{\langle \delta T_{\omega}\rangle}{T_0}.
\end{eqnarray} 
Here
$
F
= \displaystyle \biggl(\frac{\Delta_C-\mu}{T_0}\biggr)\exp\biggl(-\frac{\Delta_C-\mu}{T_0}\biggr)$,
$j^{max}$ is the maximal value of the current density from the GC to the MG
via the b-P BL, $T_0$ is the lattice temperature, $\langle \delta T_{\omega}\rangle$ is the averaged
electron effective temperature variation, and $H$ is the GC  width.
The maximal electron current density   $j^{\max}$ is given by $j^{max}= e\Sigma/\tau_{\bot}$, where $\Sigma \simeq \Sigma_D$ is  the electron density in the GC, $\tau_{\bot}$ is the electron
try-to-escape time, $\mu \simeq \mu_D$ is the electron Fermi energy, and $e$ is the electron charge.

\section{Electron energy balance in the GC and ac effective temperature }

The spatial distribution (along the axis $x$) 
of
$ \langle\delta T_{\omega}\rangle$, is described by  the one-dimensional 
electron heat transport equation, which for the device structure under consideration is presented in the following form~\cite{15}:
\begin{eqnarray}\label{eq2}
-h\frac{d^2 \langle\delta T_{\omega}\rangle}{d x^2}+ 
 \frac{\langle\delta T_{\omega}\rangle}{\tau_{\varepsilon}}
= \frac{\langle Q_{\omega}\rangle}{\Sigma}.
\end{eqnarray}
Here 
$h \simeq v_W^2/2\nu$ is the  electron thermal conductivity in the GC per electron
(this  corresponds to the Wiedemann-Franz relation),   $\tau_{\varepsilon}$ is the electron energy relaxation time in the GC, 
$\langle Q_{\omega} \rangle ={\mathrm {Re}}~\sigma_{\omega}
\langle|\delta E_{\omega}|^2\rangle $ is the Joule power (per GC area unit)  associated with
the signal  electric field, $\delta E_{\omega}$, along the GC caused by the signal voltage at the side contacts and and amplified by the plasmonic effect,
 ${\mathrm {Re}}~\sigma_{\omega} = \sigma_0\nu^2/(\nu^2+\omega^2)$, $\sigma_0 = e^2\mu/\pi\hbar^2\nu$ are the Drude ac and dc conductivities, and $\nu$ is the electron scattering frequency in the GC.
Due to the smallness of the collector window length, we disregard the energy flux of the electrons 
transferring from the GC into the MG via this  window (i.e.,the Peltier cooling)~\cite{9,15,16}
 in comparison with the electron energy relaxation on phonons in the whole GC~\cite{17,18,19,20,21} and the heat absorbed by the side contacts 
(due to a high electron heat conductivity in GCs~\cite{22,23}).

The quantity
 $\langle|\delta E_{\omega}|^2\rangle$  is the square of  ac electric field modulus  averaged over the oscillations.
Setting the boundary condition of the ac potential $\delta \varphi$ as
\begin{eqnarray}\label{eq3}
\langle\delta \varphi_{\omega}\rangle|_{x = \pm L} =\delta V_{\omega},
\end{eqnarray} 
and accounting for the excitation of plasmonic oscillations in the gated two-dimensional electron systems (neglecting the role of the short GC b-P region) 
one can obtain (see Appendix A) 
\begin{eqnarray}\label{eq4}
\langle|\delta E_{\omega}|^2\rangle = \frac{1}{2}\biggl(\frac{\delta V_{\omega}}{L}\biggr)^2\biggl|\frac{\gamma_{\omega}\sin(\gamma_{\omega}x/L)}{\cos \gamma_{\omega}}\biggr|^2.
\end{eqnarray} 
Here   $\gamma_{\omega} =\pi\sqrt{\omega(\omega+i\nu)}/2\Omega_P$ and
$\Omega_P=(\pi\,e/\hbar\,L)\sqrt{\mu\,W/\kappa}$ are the effective wavenumber and the plasmonic frequency (corresponding to the symmetric conditions at the contacts), respectively,
$\kappa$ is  the dielectric constant of the BLs, $W$ is the GL thicknesses.
The second term in the left-hand side of Eq.~(2) is associated with the electron energy relaxation due to the  interaction with phonons (mainly with the GC and interface  phonons~\cite{17,18,19,20,21}), whereas the term in the right-hand side describes the Joule heating by the ac electric field induced by the potential variations at the side contacts (i.e., by the THz radiation).

Solving Eq.~(2) with the boundary conditions
\begin{eqnarray}\label{eq5}
\langle\delta T_{\omega}\rangle|_{x = \pm L} =0,
\end{eqnarray} 
we obtain the following spatial distribution:
\begin{widetext}
\begin{eqnarray}\label{eq6}
\langle\delta T_{\omega}\rangle=  \biggl(\frac{v_W^2\tau_{\varepsilon}}{\mu\nu}\biggr)
\biggl\{
\frac{1 - \displaystyle\frac{\cos(\pi\,\omega\,x/\Omega_PL)}{1+(\pi æ\,\omega/\Omega_P)^2}
- \displaystyle\biggl[1-\frac{\cos(\pi\,\omega/\Omega_P)}{1 + (\pi æ\,\omega/\Omega_P)^2}\biggr]
\frac{\cosh(x/L æ)}{\cosh(1/æ)}
}{\sin^2(\pi\,\omega/2\Omega_P)+ (4\Omega_P/\pi\nu)^2\cos^2(\pi\,\omega/2\Omega_P)}\biggr\}\biggl(\frac{e\delta V_{\omega}}{L}\biggr)^2.
\end{eqnarray}
\end{widetext}

Figure~3  shows the spatial distributions of the temperature signal component $\langle\delta T_{\omega}\rangle$ obtained using Eq.~(6) for different ratios, $\omega/\Omega_P$, of the signal frequency and the plasmonic frequency.
It is seen that the deviation of the signal frequency from the plasmonic frequency leads to
a marked weakening of the electron heating.
The  dependences of the  $\langle\delta T_{\Omega_P}\rangle|_{x=0}$ (in the GC center) versus the GC length $2L$ for  different values of the electron collision frequency $\nu$ are plotted in Fig.~4.
The behavior of these plots is associated with the specifics of the ac electric field distribution
 (this field is a relatively small in the GC center leading to a decrease of the average field with increasing $L$, and the dependence of the electron heat conductivity $h$ on $\nu$).  
As follows from Eq.~(6), $\langle\delta T_{\Omega_P}\rangle|_{x=0}\propto{\overline \Pi}_{\Omega_P}/\nu $,  where  
 \begin{eqnarray}\label{eq7}
{\overline \Pi}_{\omega}= 1 - \displaystyle\frac{1}{1+(\pi æ\,\omega/\Omega_P)^2}\nonumber\\
- \displaystyle\biggl[1-\frac{\cos(\pi\,\omega/\Omega_P)}{1 + (\pi æ\,\omega/\Omega_P)^2}\biggr]
\frac{1}{\cosh(1/æ)}
\end{eqnarray} 
and the parameter $æ = \sqrt{h\tau_{\varepsilon}}/L = v_W\sqrt{\tau_{\varepsilon}/2\nu}/L$
characterize  the electron system  cooling at the side contacts. The factor
${\overline \Pi}_{\omega}$ and, hence,  ${\overline \Pi}_{\Omega_P}$ exhibit a pronounced variation with varying  $\nu$ and $L$ (via the dependence of  $æ$ on $\nu$ and $L$), which leads to the features of Fig.~4.

\begin{table*}[t]
\centering
\caption{\label{table} Parameters of the GC-FET  detectors with h-BN/b-AP/h-BN gate  BL and Al MG.} \vspace{2 mm}
\begin{tabular}{|r|c|c|c|c|c|c|c|c|c|c|}
\hline
  $\Delta_C$(meV)& $\Delta_M$(meV)& $\mu_D$~(meV)&$L$ ($\mu$m)&$L_C$~($\mu$m)& W (nm)  &$\Omega_P/2\pi$ (THz)& $\nu^{-1}$ (ps)& $\tau_{\varepsilon}$~(ps) &
$\tau_{\bot}$~(ps)\\ 
\hline
225 &	85& 140 &1.0&0.05 - 0.15&32 - 72 &1.0 - 1.5& 0.25 - 1.0 & 10 &10\\ 
\hline
\end{tabular}
\end{table*}

\section{Detector responsivity }

Equations~(1) and (6) result in
\begin{eqnarray}\label{eq8}
\frac{ \langle\delta J_{\omega}\rangle}{2HL_C} =
\frac{ e\mu}{\pi\hbar^2\nu\,T_0}\biggl(\frac{\tau_{\varepsilon} F}{\tau_{\bot}}\biggr)
\biggl(\frac{e\delta V_{\omega}}{L}\biggr)^2
\nonumber\\
\times
\frac{{\overline \Pi}_{\omega}}
{[\sin^2(\pi\,\omega/2\Omega_P)+ (4\Omega_P/\pi\nu)^2\cos^2(\pi\,\omega/2\Omega_P)]}.
\end{eqnarray}
We use the following definition of the  GC-FET bolometric detector voltage responsivity:
$R_{\omega}^V =  \langle\delta J_{\omega}\rangle \rho_L/S_{\omega}$, where
$\rho_L $ is the load resistance, and $S_{\omega}$
is the THz power collected by the detector antenna.

The dark current is given by~\cite{15}
\begin{eqnarray}\label{eq9}
J^{dark} = \frac{2HL_C}{\pi}\biggl(\frac{\mu}{\hbar\,v_W}\biggr)^2\exp\biggl(-\frac{\Delta_M}{T_0}\biggr)\biggl(\frac{e}{\tau_{\bot}}\biggr)\nonumber\\
\times\exp\biggl(\frac{\mu_0}{(\mu_0+\mu_D)}\frac{eV_G}{T_0}\biggr)\biggl[1-\exp\biggl(-\frac{eV_G}{T_0}\biggr)\biggr]\nonumber\\
\simeq \frac{2HL_C}{\pi}\biggl(\frac{\mu_D}{\hbar\,v_W}\biggr)^2\exp\biggl(-\frac{\Delta_M}{T_0}\biggr)\biggl(\frac{e^2V_{G}}{\tau_{\bot}T_0}\biggr)
,
\end{eqnarray}
where  $\mu_0 =\kappa\hbar^2v_W^2/4e^2W$. Here we accounted for that 
at $\mu \simeq \mu_D +\mu_0eV_G/(\mu_0+\mu_D)$. 
Considering Eq.~(9), we find that 
 at $eV_{G} \simeq T_0 \ll e{\overline V_G}$ (where ${\overline V_G} =  T_0 (\mu_0+\mu_D)/e\mu_0$),

\begin{eqnarray}\label{eq10}
\rho_L = \rho \simeq \frac{\pi}{2HL_C}\exp\biggl(\frac{\Delta_M}{T_0}\biggr)\biggl(\frac{\hbar\,v_W}{\mu_D}\biggr)^2\biggl(\frac{\tau_{\bot}T_0}{e^2}\biggr).
\end{eqnarray}
Accounting for that the antenna aperture $A_{\omega} =\lambda_{\omega}^2g/4\pi$, 
where $\lambda_{\omega}$ is the THz radiation wavelength and $g$ is the antenna gain~\cite{24}, for the relation between $|\delta V_{\omega}|^2$ and  the collected power we obtain $|\delta V_{\omega}|^2 = 32S_{\omega}/gc$, where $c$ is the speed of light in vacuum.
Using the latter and invoking Eq.~(7), we obtain

\begin{eqnarray}\label{11}
R_{\omega}^V= \frac{R^V{\overline \Pi}_{\omega}}
{\sin^2(\pi\,\omega/2\Omega_P)+ (4\Omega_P/\pi\nu)^2\cos^2(\pi\,\omega/2\Omega_P)},
\end{eqnarray}
where 

\begin{eqnarray}\label{eq12}
R_V = 
\frac{32}{137g}\biggl(\frac{\hbar}{eT_0}\biggr)
\biggl(\frac{\Delta_M}{\mu}\biggr)
\biggl(\frac{v_W^2\tau_{\varepsilon}}{L^2\nu}\biggr)
\nonumber\\
=\frac{32}{137g}\biggl(\frac{\hbar\Delta_M}{e^2T_0}\biggr)
\biggl(\frac{\tau_{\varepsilon}{\mathcal M}}{L^2}\biggr)
\end{eqnarray}
is the responsivity characteristic value. Here ${\mathcal M}$ is the electron mobility in the GC. 
At $\nu < \Omega_P $ the  denominator in the right-hand side describes  the resonant frequency
dependence of  $R_{\omega}^V$ [with maxima of the latter at $\omega \simeq \Omega_P(2n-1)$, $n = 1,\,2,\,3,...$ is the resonance index].
The quantity $R_{\Omega_P}^V = R^V {\overline\Pi}_{\Omega_P}$ is equal to the maximal voltage responsivity at the fundamental resonance $\omega = \Omega_P$ [see Eq.~(7)].
It is instructive that $R^V$ is independent of the length of the collecting window $2L_C$ and the try-to-escape time $\tau_{\bot}$.

Using the device structure parameters for the  detector structure with $\nu =$~2/ps, $L_C = 0.1~\mu$m, $H = 2~\mu$m, and  $g = 1.6$ 
(other parameters are listed in Table I), 
and accounting for  Eqs.~(9) - (11), for the resonant responsivity we obtain
$R_{\Omega_P}^V \simeq 7\times 10^3$~V/W.

\begin{figure}[t]
\centering
\includegraphics[width=8.0cm]{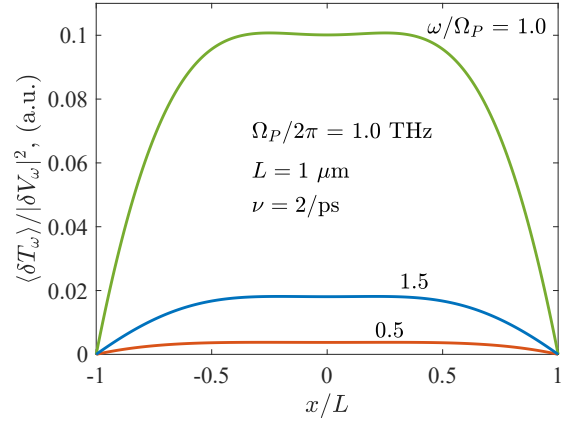}
\caption{Spatial distribution of the  electron temperature variation
caused by THz radiation with different frequencies. 
}
\label{Fig3}
\end{figure}

\begin{figure}[t]
\centering
\includegraphics[width=8.0cm]{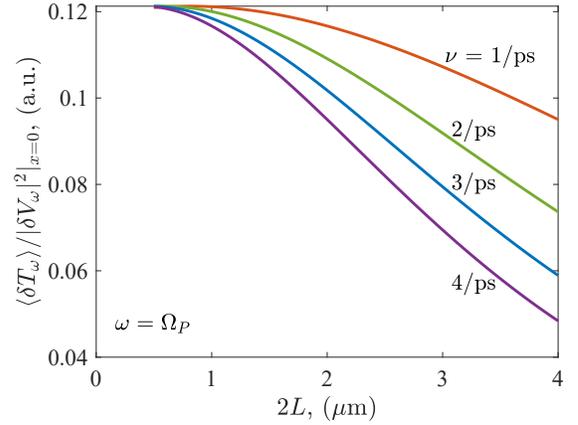}
\caption{Electron temperature variation at the GC center ($x=0$) as a function of the GL length $2L$ calculated for different values of the electron collision frequency $\nu$
(different values of the electron heat conductivity $h$: 5, 5/2, 5/3, 5/4 $\times 10^3$~cm$^2$/s) at $\omega = \Omega_P$.
}
\label{Fig4}
\end{figure}

\begin{figure}[t]
\centering
\includegraphics[width=8.0cm]{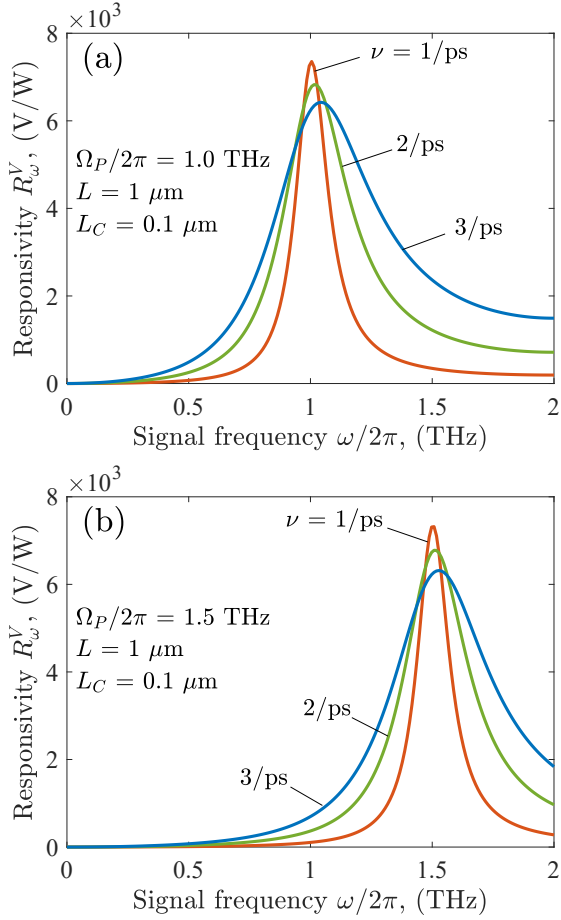}
\caption{Spectral characteristics  of the voltage responsivity  $R_{\omega}^V$ (in the frequency range around the fundamental resonance) for different values of collision frequency $\nu$ calculated for GC-FETs with different plasmonic frequencies (a) $\Omega_p/2\pi = 1$~THz and (b)  $\Omega_p/2\pi = 1.5$~THz.
}
\label{Fig5}
\end{figure}

\begin{figure}[t]
\centering
\includegraphics[width=8.0cm]{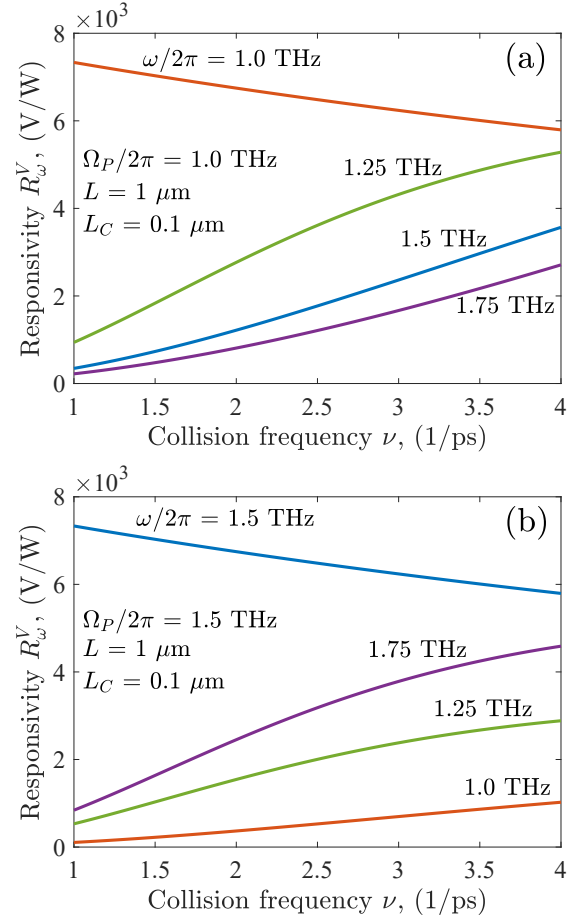}
\caption{Voltage responsivity $R_{\omega}^V$ versus collision frequency $\nu$
at different signal frequencies $\omega/2\pi$ calculated for the GC-FETs with (a) 
$\Omega_p/2\pi = 1$~THz and (b) $\Omega_p/2\pi = 1.5$~THz. The top lines correspond to the fundamental plasmonic resonances.
}
\label{Fig6}
\end{figure}

\section{Detector noise equivalent power and dark current limited detectivity}

\begin{figure}[t]
\centering
\includegraphics[width=8.0cm]{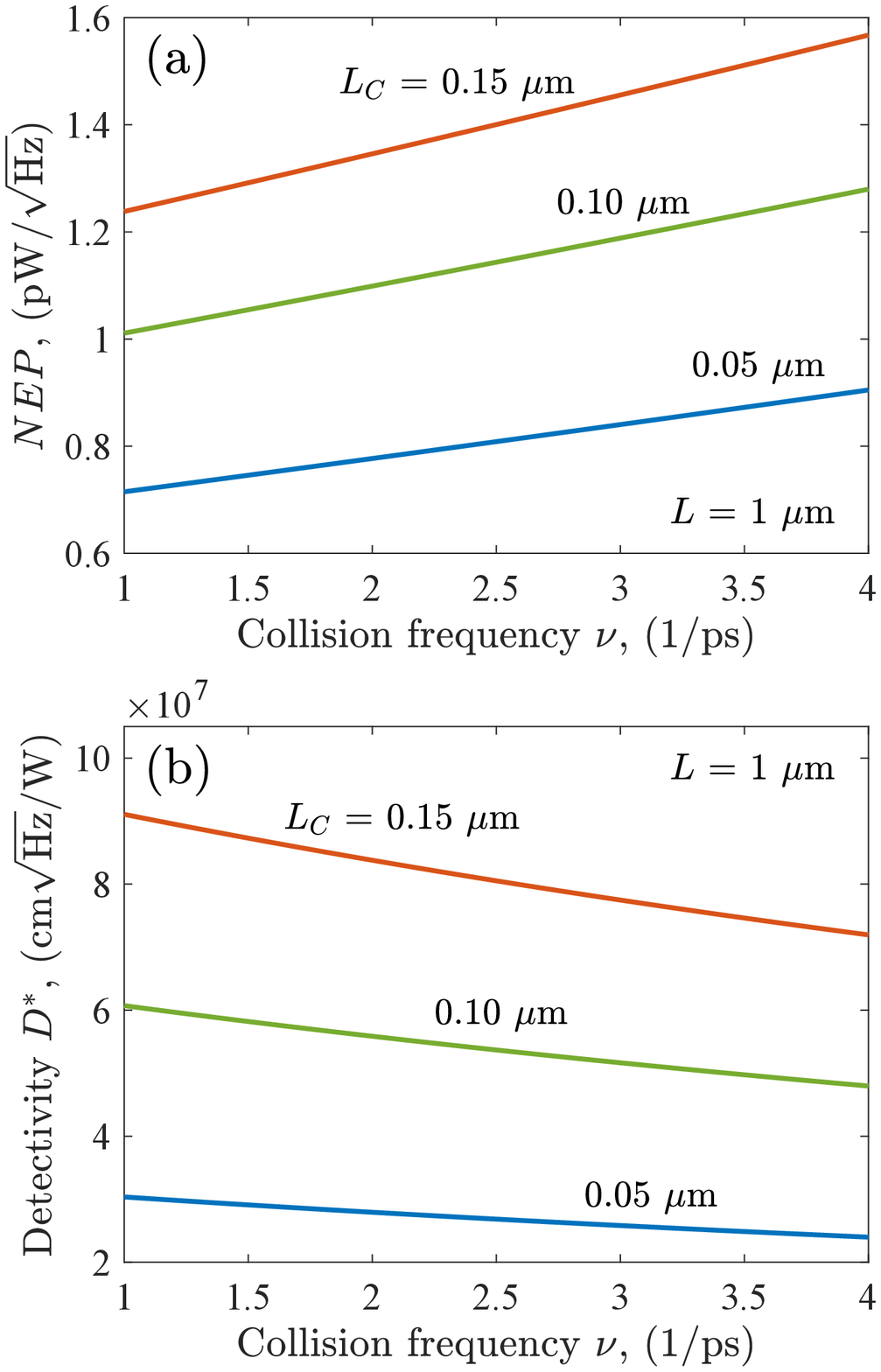}
\caption{Dependencies of (a) noise equivalent power ($NEP$) and (b) dark current limited detectivity $D^*$ on the collision frequency $\nu$ calculated for the GC-FETs with different lengths $L_C$ of the b-P collector window BL at the resonant frequency ($\omega =\Omega_P$).
}
\label{Fig7}
\end{figure}

\begin{figure}[t]
\centering
\includegraphics[width=8.0cm]{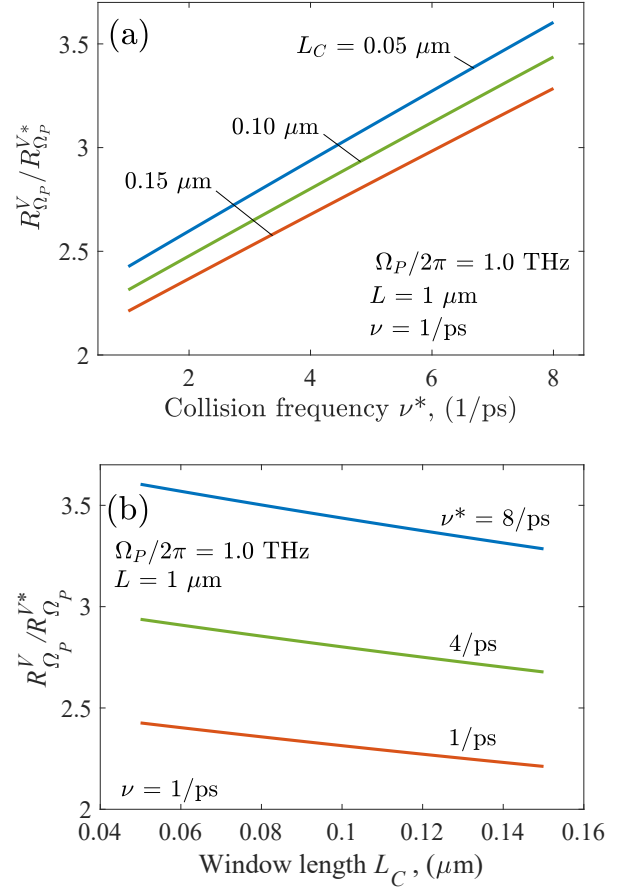}
\caption{Ratio of the  responsivities, 
$R_{\Omega_P}^V/R_{\Omega_P}^{V^*}$,  of  GC-FET detectors with the composite h-BN/b-P/h-BN gate BL and with the b-P gate BL (a) as a function of the electron collision frequency  $\nu^*$
for different  $L_C$ 
and (b) as a function of  $L_C$  for different $\nu^*$ (at $\nu = $~1/ps
and $L = 1.0~\mu$m).
}
\label{Fig8}
\end{figure}

Using Eqs. (9) and (11), we arrive at the following expressions for the GC-FET noise equivalent power (NEP)
and the dark-current-limited detectivity ($D^*$) at the fundamental plasmonic resonance ($n=1$):

\begin{eqnarray}\label{eq13}
{\mathrm {NEP}} \simeq \frac{137\sqrt{2\pi} gL^2}{64\sqrt{2L_CH}}
\biggl(\frac{\sqrt{\tau_{\bot}}\nu\,T_0}{v_W\tau_{\varepsilon}}\biggr)\nonumber\\
\times \biggl(\frac{2T_0}{\Delta_M}\biggr)
\exp\biggl(\frac{\Delta_M}{2T_0}\biggr)\frac{1}{{\overline \Pi}_{\Omega_P}},
\end{eqnarray}
\begin{eqnarray}\label{eq14}
D^* = \frac{64(2L_CH)}{137\sqrt{2\pi}\,gL^2}\biggl(\frac{v_W\tau_{\varepsilon}}{\sqrt{\tau_{\bot}}\nu\,T_0}\biggr)\nonumber\\
\times
\biggl(\frac{\Delta_M}{2T_0}\biggr)\exp\biggl(-\frac{\Delta_M}{2T_0}\biggr){\overline \Pi}_{\Omega_P},
\end{eqnarray}
respectively, where
$$
{\overline \Pi}_{\Omega_P} = 1-\frac{1}{1+(\pi æ)^2} - \biggl[1+\frac{1}{1+(\pi æ)^2}\biggr]\frac{1}{\cosh(1/æ)}.
$$

The estimates for $NEP$ and $D^*$ of the GC-FET detector with the same parameters
as those used for the estimation of $R_{\Omega_P}^V$ yield
$NEP   \sim  1$~pW/$\sqrt{{\mathrm {Hz}}}$ and
$D^* \sim 6\times 10^7$~cm$\sqrt{{\mathrm {Hz}}}$/W.
These values are about or exceeding the characteristics  of other uncooled
THz bolometers (see, for example, Refs.~\cite{25} and \cite{26}).

Figure~7 shows $NEP$ and $D^*$ as functions of $\nu$ calculated for the GC-FETs with different lengths, $L_C$, of the collector window.



\section{Comments}

Calculating Eq.~(9), we disregarded the variation of $\mu$ with increasing bias voltage $V_G$ 
at not too large bias voltages, $\mu \simeq \mu_D +\mu_0 eV_G/(\mu_0+\mu_D) \simeq \mu_D$. 
For $\kappa = 4 $ and the same other parameters as above, $\mu_0 \simeq (6.8 - 13.6)$~meV and  
${\overline V_G} \simeq (283 - 541)$~meV. 
Hence, at $eV_G \lesssim  T_0$, one obtains 
 $(R_{\omega}^V/R_{\omega}^V|_{V_G = 0} -1) \simeq (4.6 - 8.8)~\%$.

According to Eq.~(9), the dark current $J^{dark} = eV_G/\rho$, where  $\rho$ is given by Eq.~(10).
Taking this into account, in particular,  for the ratio
of $D^*$ in a wide range of $V_G$  and $D^*$ at $eV_G = T_0$
given by Eq.~(14) we obtain
$D^*/D^*|_{eV_G = T_0} \simeq \sqrt{T_0/eV_G}$. It is worth noting that
at smaller gate voltages $eV_G < T_0$, the detector detectivity might be limited not by the dark current, but by the Johnson-Nyquist (J-N) noise (see, for example,~Ref.~\citep{27}). 
Indeed, the J-N noise is $i_{i,J-N}^2 = (4T_0/\rho)\delta f$, while the dark current noise is $i_{n,dark}^2 = 4eJ^{dark}\delta f = (eV_G/\rho)\delta f$,
for the ratio of the pertinent detectivity values we obtain $D^*/D^*_{J-N} \simeq \sqrt{T_0/eV_G}$. 
However, the optimum value of the gate bias would also depend on the exposure level and the readout circuitry (see Ref.~\citep{28} for more details).

The present study does not include the consideration of high bias voltages since, in this case, the
"thermionic model" 
becomes invalid 
and the electron tunneling from the GC through the b-P BL must be accounted for.

The parameters assumed above correspond to the electron mobility in the GC in the range of 
${\mathcal M} \simeq (1.78 - 7.14)\times 10^4$~cm$^2$/V$\cdot$s.
For the chosen electron density, these values for the room temperature mobility in the GCs, a substantial part of which is  encapsulated by b-BN, are fairly realistic (see, for example,~Refs.~\cite{11,12,13}).
The elevated values of the GC  room-temperature mobility and, hence, decreased electron collision frequencies, can be also achieved in the GC-FET structures with the sections of the WSe$_2$ gate BL~\cite{29} replacing the pertinent  h-BN sections. 
Similar GC-FET bolometric detectors can comprise the b-AsP or b-As central BL sections.

Below we compare the voltage responsivities, $R_{\Omega_P}^V$ and $R_{\Omega_P}^{V*}$, 
of the GC-FET detectors with the composite h-BN/b-P/h-BN gate BL 
and the uniform b-P gate BL~\cite{15}  at $\omega=\Omega_p$, assuming that the load
resistances are equal to their terminal resistances for both detectors.
For their responsivities, we obtain the plots in Fig.~8  (the asterisk marks the responsivity of  the GC-FETs with the b-P gate BL). 
Figure~8(a) shows the ratio of the responsivities $R_{\Omega_P}^V/R_{\Omega_P}^{V*}$, 
as a function of the collision frequency, $\nu^*$, in the GC-FETs with the b-P gate BL,
for  the the collision frequency in the FET with the composite BL equal to $\nu = $~1/ps and different  b-P window BL length values $L_C$.
Figure~8(b) shows the dependence of these responsivities on $L_C$ calculated for different $\nu^*$. 
As seen in Fig.~8, the incorporation of the composite gate BL might provide a marked enhancement of the GC-FET detector performance.
The fact that the ratio $R_{\Omega_P}^V/R_{\Omega_P}^{V*}$ increases with increasing $\nu^*/\nu$ relatively slowly  is attributed to
a weakening of the electron cooling at the side contacts when $\nu^*$ rises.

When electron collision frequency in  the GC central  (collector window) section $\nu_C \gg \nu$, the contribution of this section to the Joule heating can be marked. This might limit the applicability of the model used above to a relatively small $L_C/L$.

Now we compare the net Joule power produced in the side sections, $2L\overline{ \langle Q_{\omega}\rangle}$, and in the central section, $2L_C\overline {\langle Q_{\omega_C}\rangle}$, at the fundamental plasmonic resonance $\omega = \Omega_P$ for the limiting case $\nu \ll \Omega_P \lesssim \nu_C $. 
Equations~(A3) and (A4) (see Appendix A) give rise to the following estimate of the ratio:

\begin{eqnarray}\label{eq15}
\frac{L\overline {\langle Q_{\omega}\rangle}}{L_C\overline { \langle Q_{\omega,C}\rangle}} \simeq
\frac{\nu^2}{\nu_C^2}\sqrt{\frac{\nu_C^2 +\omega^2}{\nu^2+\omega^2}}
\nonumber\\
\times \biggl|\frac{\cos(\gamma_{\omega,C}L_C/L)}{\cos(\gamma_{\omega}L_C/L)}\biggr|^2
\frac{\int_{-L}^{L}dx |\sin(\gamma_{\omega}x/L)|^2}
{\int_{-L_C}^{L_C}dx |\sin(\gamma_{\omega,C}x/L)|^2}.
\end{eqnarray}
At the fundamental plasmonic resonance ($\omega = \Omega_P$), assuming that
 $\nu \ll \Omega_P \lesssim \nu_C$,  Eq.~(15) yields the estimate:

\begin{eqnarray}\label{eq16}
\frac{L\overline {\langle Q_{\omega}\rangle}}{L_C\overline { \langle Q_{\omega,C}\rangle}} \simeq \biggl(\frac{\nu}{\nu_C}\biggr)^2 \biggl(\frac{L}{L_C}\biggr)^3.
\end{eqnarray}
This implies that 
 the simplification made in Sec.~IV  is justified when $L_C/L \ll (\nu/\nu_C)^{2/3} \simeq (M_C/M)^{2/3}$. Here $M$ and $M_C$ are the electron mobilities in the pertinent sections of the GC. In the case of the h-BN/b-P/h-BN composite BL at room temperature, one can set $M_C \simeq 8\times 10^3$~cm$^2$/Vs~\cite{10},  and 
 $M \simeq(3 - 6)\times 10^4$~cm$^2$/Vs~\cite{11} (i.e., $\nu \simeq (2- 3)$/ps as was used for the dependences in Figs.~3 - 5).
The latter values correspond to the condition $L_C/L < 0.26 - 0.41 $, which was satisfied in the above  calculations. 
At a small $L_C/L$ and a large $\nu_C$, the electron heating in the middle section can be not too small. 
However, this circumstance can be  
 beneficial leading to the enhancement of  the GC-FET bolometric detectors under consideration.
 
 Our device model intended for the GC-FET bolometric detectors with the composite BL assumes that the effect of Peltier cooling is weaker than the electron energy
 relaxation due to the interaction with phonons.
 The pertinent condition means $(L_C/L)(\tau_{\varepsilon}/\tau_{\bot})(\Delta_C\Delta_M/T_0^2)\exp(-\Delta_M/T_0)\ll 1$. Using the parameters listed in Table I,
we arrive at the following criterion $L_C/L \ll \tau_{\bot}/\tau_{\varepsilon}$.
If $\tau_{\bot} \lesssim \tau_{\varepsilon}$, this condition is not burdensome.

The effect of the electron thermal transport along the GC and of the side contacts cooling can be diminished in the GC structures shown in Fig.~1(c). 
In the structure with a short b-P window BL and a short MG shown in Fig.~1(c), the spacing $2L$ can be  relatively large without sacrificing the plasmonic frequency  $\Omega_P$.
This is because the ungated GCs can exhibit rather large  $\Omega_P$ even at relatively  long $2L$.
The device model of this GC-FET bolometric detector is more complex and requires a separate study.

\section*{Conclusions}
We predicted that implementation of the GC-FET bolometric detectors with the composite h-BN/b-P/h-BN gate BL   having a narrow b-P collector window could substantially enhance the detector performance in comparison with the GC-FETs with the uniform b-P gate BL, especially invoking the plasmonic resonances.

\section*{Author's contributions}
All authors contributed equally to this work.

\section*{Acknowledgments}

The work at RIEC, UoA, and UB was supported by the Japan Society for Promotion of Science (KAKENHI  Nos. 21H04546, 20K20349),
Japan and the RIEC Nation-Wide Collaborative research
Project No. R04/A10, Japan.  The work at RPI was supported by AFOSR (contract number FA9550-19-1-0355).

\section*{Conflict of Interest}

The authors declare no conflict of interest.

\section*{Data availability}
All data that support the findings of this study are available within the article.

\section*{Appendix A.} 
\setcounter{equation}{0}
\renewcommand{\theequation} {A\arabic{equation}}

The signal component of the potential $\ \delta\varphi_{\omega}$
in the gated GC is governed by the following equations~\cite{6,15}:
\begin{eqnarray}\label{eqA1}
\frac{d^2\delta\varphi_{\omega}}{d x^2} + \gamma_{\omega}^2\delta\varphi_{\omega} =0
\end{eqnarray}
at $L_C < |x| <L$, and
\begin{eqnarray}\label{eqA2}
\frac{d^2 \delta\varphi_{\omega}}{d x^2} + \gamma_{\omega,C}^2\delta\varphi_{\omega} =0
\end{eqnarray}
at $ 0\leq |x| <L_C$. Here 
$\gamma_{\omega} = \pi\sqrt{\omega(\omega + i\nu)}/2\Omega_P$, $\gamma_{\omega,C} = \pi\sqrt{\omega(\omega + i\nu_C)}/2\Omega_P$, and
$\nu$ and $\nu_C$ are the electron collision frequencies in the pertinent section of the GC.
Apart from the boundary conditions~(3), these equations are supplemented by the conditions of the potential continuity at $x = \pm L_C$.

In the case $L_C \ll L$, neglecting the contribution of a narrow region 
$-L_C < x < L_C$, i.e., extending the validity of Eq.~(A1) all over the whole GC,
we obtain
\begin{eqnarray}\label{eqA3}
 \delta\varphi_{\omega}=\delta V_{\omega}\frac{\cos(\gamma_{\omega}x/L)}{\cos\gamma_{\omega}}.
\end{eqnarray}
This equation leads to Eq.~(4) in the main text.

The electron collision frequency in the b-P collector window region $\nu_C$ differs from that, $\nu$, in the main part of the GC-FET structure, where it is determined by the GC-n-BN interfaces.
In the devices under consideration, the ratio $\nu_C/\nu$ can be rather large.
Solving Eqs.~(A1) and (A2), we obtain the spatial potential distribution
in the region  $L_C <|x| < L$ coinciding with with that given by Eq.~(A3), while the distribution in the region $-L_C <x < L_C $ is given by 
\begin{eqnarray}\label{eqA4}
 \delta\varphi_{\omega}=\delta V_{\omega}\frac{\cos(\gamma_{\omega,C}x/L)}
 {\cos\gamma_{\omega}}\frac{\cos(\gamma_{\omega}L_C/L)}
 {\cos(\gamma_{\omega,C}L_C/L)}.
\end{eqnarray}

If $\nu_C = \nu$, Eqs.~(A3) and (A4) coincide, and Eqs.~(4) -(12) are valid for
a wide range of ratio $L_C/L$.

\end{document}